\documentclass[12pt,a4paper]{article}
\pdfoutput=1 
\usepackage[utf8]{inputenc}
\usepackage{csquotes}
\usepackage{graphicx}
\usepackage{hyperref}
\usepackage[margin=2cm]{geometry}

\usepackage[textsize=footnotesize]{todonotes}
\usepackage{pifont}
\usepackage{amssymb}
\usepackage[natbib, style=apa]{biblatex}

\title{A framework for digital trace data collection through data donation}
\author{L. Boeschoten, J. Ausloos, J.E. M\"{o}ller, T. Araujo \& D.L. Oberski}

\begin{document}
\maketitle

\begin{abstract}
A potentially powerful method of social-scientific data collection and investigation has been created by an unexpected institution: the law. Article 15 of the EU's 2018 General Data Protection Regulation (GDPR) mandates that individuals have electronic access to a copy of their personal data, and all major digital platforms now comply with this law by providing users with ``data download packages'' (DDPs). Through voluntary donation of DDPs, all data collected by public and private entities during the course of citizens' digital life can be obtained and analyzed to answer  social-scientific questions -- with  consent. Thus, consented DDPs open the way for vast new research opportunities. However, while this entirely new method of data collection will undoubtedly gain popularity in the coming years, it also comes with its own questions of representativeness and measurement quality, which are often evaluated systematically by means of an error framework. Therefore, in this paper we provide a blueprint for digital trace data collection using DDPs, and devise a ``total error framework" for such projects. Our error framework for digital trace data collection through data donation is intended to facilitate high quality social-scientific investigations using DDPs while critically reflecting its unique methodological challenges and sources of error. In addition, we provide a quality control checklist to guide researchers in leveraging the vast opportunities afforded by this new mode of investigation.
\end{abstract}

\section{Introduction}

Digital traces left by citizens during the natural course of modern life hold an enormous potential for social-scientific discoveries \citep{kingEnsuringDataRichFuture2011}, because they can measure aspects of our social life that are difficult or impossible to measure by more traditional means \citep{pentland2010honest}. For example, classic sociological theory describes citizens' interactions \citep{colemanFoundationsSocialTheory2000}, but large-scale data on such interactions are only now becoming available from digital platforms \citep[e.g.][]{szellMultirelationalOrganizationLargescale2010}. Similarly, experiments show that news reports can produce different opinions depending on the consumer's political motivations \citep[e.g.][]{bolsenInfluencePartisanMotivated2014}, but only through digital media we can now  observe the simultaneous dynamics of consumed (mis)information, motivation, and opinion. With increased datafication and digitalization of our societies, the study of digital traces gains even more relevance. As more and more of our social lives happens on platforms, the digital traces we leave behind on those platforms become an important object of study \citep{papacharissi2010networked}. Further examples of digital traces' potential abound, and indeed, digital trace data collected through Application Programming Interfaces (APIs) and web scraping have been used in many applications, including network analysis from mobile phone data \citep{blondelSurveyResultsMobile2015}; price indexing from online shop prices \citep{de2013online}; political opinion and electoral success prediction from Twitter data \citep{schoenPowerPredictionSocial2013, jungherrAnalyzingPoliticalCommunication2015}; and personality profiling from Facebook ``likes" (\cite{kosinskiPrivateTraitsAttributes2013}; see also \cite{settanniPredictingIndividualCharacteristics2018} for an overview of similar studies). 

In recent times, however, the faucet of social science data from APIs and web scraping has been decisively turned off by the relevant tech companies \citep{bruns2019after, perriam2020digital, freelon2018computational}. Through mutual agreement and negotiation between academia and industry, new efforts to make such data available to social scientists are underway, for example through ``Social Science One'' \citep{kingNewModelIndustry2019}. These data are now becoming available in aggregated form under strict privacy protections \citep{dorazioDifferentialPrivacySocial2015, messingFacebookPrivacyProtectedFull2020}. While this new collaborative model is useful for social-scientific investigation of certain research questions, it does not fit all purposes mentioned above and has raised concerns about the role of platforms (for an overview, see \citet{halavais2019overcoming}). First, by their very definition, the imposed data protection regulations ensure these data cannot address questions of individual (user-level) dynamics or networks \citep{Oberski2020Differential}. Second, APIs provide public data only; much of digital trace data's putative power, however, lies in private data that is too sensitive to share, such as location history, browsing history, or private messaging \citep{quan2010uses}. 
Third, the available data generally pertain to a nonrandom subset of the digital platform's user group (e.g. Facebook or Twitter) which is not representative of many populations of social-scientific interest \citep{mellonTwitterFacebookAre2017, pfeffer2018tampering}. Fourth, for both approaches, the researcher is entirely dependent on the private company that holds the data; sudden retractions of this collaborative spirit can, and have, occurred, posing a risk to the research process \citep{bruns2019after}. In addition. there is no possibility to independently verify that the data is complete and checked for measurement errors. Finally, even when a data processing company decides to share data for scientific purposes, the citizens who actually generated those data are generally impossible to contact for their consent, in some cases putting a firm legal basis for further data analysis in question (for example, following article 6, EU General Data Protection Regulation, or similar laws in other jurisdictions). Issues two and three regarding private data and nonrandom subsets can be overcome by collecting data via the installation of a tracker, plugin or app, such as performed by \citet{andrews2015beyond, reeves2019screenomics, haenschen2020self, araujo2017much} and \citet{scharkow2019reliability}.

\begin{figure}[tb]
    \includegraphics[width=\textwidth]{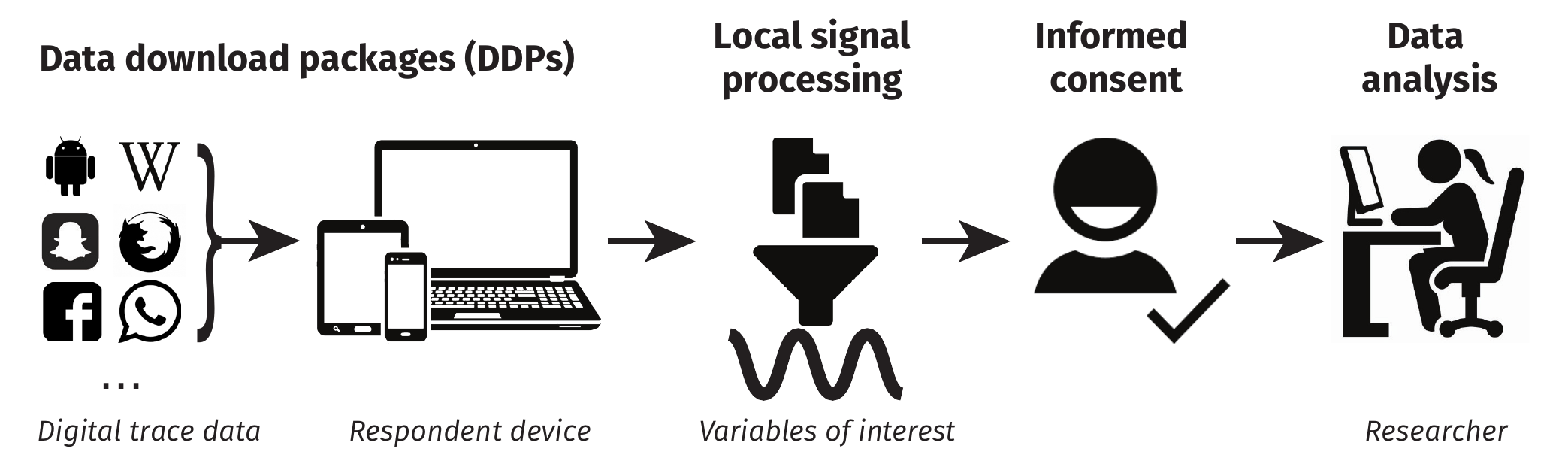}
    \caption{A workflow illustration how a respondent's data download packages (DDPs) can be leveraged for social-scientific research after local processing and informed consent.}
    \label{fig:high-level}
\end{figure}

In this paper, we present an alternative workflow to collect and analyze digital traces, based on \emph{data download packages} (DDPs). As of May 2018, any entity, public or private, that processes the personal data of citizens of the European Union is legally obligated by the EU General Data Protection Regulation \citep[GDPR][]{gdpr2016} to provide that data to the data subject upon request, and in digital format (GDPR Article 15; \cite{ausloosGDPRTransparencyResearch2019}). Most major private data processing entities, comprising social media platforms as well as smartphone systems, search engines, photo storage, e-mail, banks, energy providers, and online shops comply with this right to data access by providing DDPs to the data subjects. To our knowledge, most large companies that operate internationally provide the same service to their users outside European Union. Our proposed workflow consists of five steps (see Figure \ref{fig:high-level}). First, data subjects are recruited as respondents using standard survey sampling techniques \citep{valliantPracticalToolsDesigning2018} and the researcher determines which DDPs are relevant for the particular research question under investigation. Second, respondents request their DDPs with the various selected providers, storing these locally on their own device. Third, stored DDPs can then be locally processed to extract relevant research variables, after which consent is requested of the respondent (step four) to send these derived variables to the researcher for analysis (step five). 
To aid researchers in planning, executing, and evaluating studies that leverage the richness of DDPs, we discuss the steps involved our proposed workflow.  The short explanation of our proposed workflow already illustrates that in this process, decision should be made regarding both measurements and representation. In such cases, traditional survey research has benefited greatly from the ``total survey error'' framework  \citep{grovesTotalSurveyError2010}; here we therefore present DDP data collection in a ``total error'' framework \citep{Biemer2016errors} adapted specifically to this new mode of data collection \citep[see also][for a generic total error framework for ``big data'']{amayaTotalErrorBig2020}. 

\bigskip 
%e.g., indicating that the process exists/is important, that it is complex, and that this paper aims to aid the researcher to perform high-quality research by presenting a total error framework to guide them through the process

The aim of this paper is to introduce and discuss the idea of data donation for scientific research. As processing DDPs in such a way that high quality research can be performed is complex and challenging, a total error framework is introduced to guide researchers through this process. We first briefly discuss the right of access in the GDPR in the next section. We then present an example research question that might be addressed using Instagram DDPs collected from Dutch adolescents in the ``Adolescents, Well-being \& Social Media'' (AWeSome) study \citep{beyensEffectSocialMedia2020}. Subsequently we present our total error framework for DDPs, and discuss the steps involved in answering such a research question in the context of this framework. Finally, we discuss limitations of our approach, as well as future directions for methodological investigation. Appendix \ref{app:checklist} provides a ready-to-use checklist as a guideline for researchers evaluating or conducting DDP studies. 

\section{The right of access in the GDPR}
In recent years, jurisdictions around the world have enacted or are in the process of enacting new data protection legislation. Examples outside the EU include the 2017 Japanese Amended Act on the Protection of Personal Information (AAPI 2016), the 2020 Brazilian General Data Protection Law (LGDP 13.709/2018), the 2020 California Consumer Privacy Act (375/2018),  the 2019 New York SHIELD act (S5575B/2019), and the proposed Personal Data Protection Bill (PDP Bill 2019) in India. Many of these laws have been designed specifically for their compatibility with the European Union's wide-reaching data protection legislation \citep{sudaJapanPersonalInformation2020, singhTechnicalLookIndian2020}, the General Data Protection Regulation (GDPR), which has applied across the EU and the UK since May of 2018. Together, these jurisdictions alone comprise about 2.2 billion people, over a quarter of the world's population. 

The GDPR grants all natural persons (``data subjects''), whatever their nationality or residence, certain rights regarding their ``personal data'' with respect to ``data controllers'', such as tech companies, governments, mobile phone providers, etc. Although the GDPR is currently likely best known among data analysts for restricting what data controllers can do with personal data, the GDPR also grants data subjects the \emph{right of access} (Article 15). This  entails ``the right to obtain from the controller confirmation as to whether or not personal data concerning him or her are being processed, and, where that is the case, \textbf{access to the personal data}...'' (Article 15.1; emphasis added). Note that Article 15 also enables access to information regarding data recipients and sources, retention periods and data derived from your personal data.
Article 15.3 further specifies the obligation for controllers to provide a copy of personal data, requiring them to do so “in a commonly used electronic form” whenever the data subject made their request by electronic means. The GDPR further grants the \emph{right to data portability} in the closely related article 20, which states: ``The data subject shall have the right to receive the personal data concerning him or her, which he or she has provided to a controller, in a structured, commonly used and machine-readable format and have the right to transmit those data to another controller without hindrance from the controller to which the personal data have been provided''. 

In practice, most large ``data controllers'' currently comply with the right of access to one's personal data and the right to data portability by providing users with the option to retrieve an electronic ``data download package'' (DDP). For example, at the moment of writing, Google provides a ``takeout'' option\footnote{\url{https://support.google.com/accounts/answer/3024190}}, and Facebook\footnote{\url{https://www.facebook.com/help/1701730696756992}},  WhatsApp\footnote{\url{https://faq.whatsapp.com/general/account-and-profile/how-to-request-your-account-information/}}, Instagram\footnote{\url{https://help.instagram.com/181231772500920?helpref}}, Uber\footnote{\url{https://help.uber.com/riders/article/request-a-copy-of-your-uber-data?nodeId=2c86900d-8408-4bac-b92a-956d793acd11}}, Apple\footnote{\url{https://privacy.apple.com/}}, Netflix\footnote{\url{https://www.netflix.com/account/getmyinfo}}, and Microsoft\footnote{\url{https://support.microsoft.com/en-us/help/4468251/microsoft-account-view-your-data-on-the-privacy-dashboard}} provide similar tools. Compliance with the right of data portability has sometimes been less straightforward for other data-controllers \citep{wong2019right}. To our knowledge, with the exception of WeChat, none of the large global data controllers limit use of these tools to the European Union. Indeed, all other legislation mentioned above -- including the California Consumer Privacy Act -- grant some right of access, though often more limited than that found in the GDPR.  
Pursuant to GDPR article 20, data controllers cannot arbitrarily limit the data they provide in this package, or prevent their users from sharing its contents with third parties, such as social scientists. Third parties may freely process such packages, for example for scientific purposes, on the basis of user consent (articles 20 and 6). The right of access is limited in that it cannot be invoked to infringe on the rights or freedoms of others, particularly on other natural persons' data protection rights, or on trade secrets; thus, the provided data should not a priori include personal data pertaining to other people \citep{wachterCounterfactualExplanationsOpening2017}. For example, Facebook's data download packages do not include information on the user's ``friends'' (only the interactions these ``friends'' have with the data subject), nor does it provide details regarding Facebook's proprietary algorithms. In this sense, data included in DDPs are limited. Furthermore, in keeping with other rights granted by the GDPR, data subjects may also request deletion of their own data. 

In spite of the limitations of the right of access, a wealth of information is contained in data download packages offered as its direct consequence. At the time of writing it appears likely that a large proportion of persons globally who use a smartphone or the internet will have some data in their DDPs. In the following section, we discuss how this fact can be leveraged for novel social-scientific research, as well as the pitfalls and errors that must be controlled along the way.

\section{Using data-download packages (DDPs) for scientific research}

To illustrate the considerations relevant when using DDPs for social-scientific research and thereby showing it potential, we will take the example of one hypothetical research question that may be of interest to social scientists, and that we think could be answered using DDP collection. However, many other research questions can very well be answered by using DDP collection. For example research questions recently investigated using APIs and webscraping, such as the previously discussed network analysis from mobile phone data \citep{blondelSurveyResultsMobile2015}, price indexing from online shop prices \citep{de2013online}, political opinion and electoral success prediction from Twitter data \citep{schoenPowerPredictionSocial2013, jungherrAnalyzingPoliticalCommunication2015}, and personality profiling from Facebook ``likes" (\cite{kosinskiPrivateTraitsAttributes2013} can be investigated while being more explicit regarding the generalizability of the findings. Alternatively, research questions typically investigated using surveys, such as energy consumption \citep{guerra2010occupants}, time spent \citep{eleveltWhereYouUsing2019} or budget research \citep{breedveld2002background} can be executed without suffering from issues such as recall bias or bias due to social desirability.

Our exemplary research question is inspired by the ``Adolescents, Well-being \& Social Media'' (AWeSome) project \citep{beyensEffectSocialMedia2020}. In a first study, adolescents (ages 14--15) in Dutch classrooms, as well as their parents and teachers, were approached by the researchers to participate in a series of surveys. Study participants were asked to answer a single survey question regarding their well-being six times per day on their mobile phone for a week. In addition, at a separate occasion they were asked to complete a survey regarding their use of WhatsApp and Instagram; in this small study, it appeared that Instagram was the most widespread social network. Furthermore, the use of self-reporting of well-being in this study may be a limitation because of recall bias \citep{beyensEffectSocialMedia2020}. 

Here, we anticipate a larger follow-up study in which adolescents' emotions are investigated \emph{using} information obtained from their Instagram DDPs. For illustration purposes, we will work with a simple, descriptive, example research question:
\begin{quote}
    Example RQ: \emph{How do emotions of Dutch adolescents differ when they are at home compared to when they are not?}
\end{quote}
To answer this question, we must obtain (1) the consent and participation of a larger group of Dutch adolescents and their parents, and (2) measurements of the participants' emotions, as well as a measure of whether they are at home or not.

Here we will discuss the steps that would be required to obtain these data using DDPs. At each of these steps, errors can occur. In order to obtain useful answers to our research question, we must therefore take account of, and, where possible, control such errors. To enumerate the error sources associated with each step in a data collection, a highly convenient framework is the \emph{total error} framework \citep{japec2015big, Biemer2016errors}. In a total error framework, each step of the data collection process is described, together with the errors that might arise from that step. The final ``total'' error in the analysis or statistics produced is then a combination of the sequence of preceding errors.  The concept of ``total error'' arose from the survey methodology literature \citep{grovesTotalSurveyError2010}, where ``total survey error'' (TSE) is the standard framework for designing, evaluating, and optimizing data collection \citep{biemerIntroductionSurveyQuality2003, biemerTotalSurveyError2010}. \citet{amayaTotalErrorBig2020} extended this framework to generic ``big data'' studies, \citet{sen2019total} extended this framework to digital trace data and \citet{Beinhauer2020} extended the framework to sensor data. 

\begin{figure}[tb]
    \includegraphics[width=\textwidth]{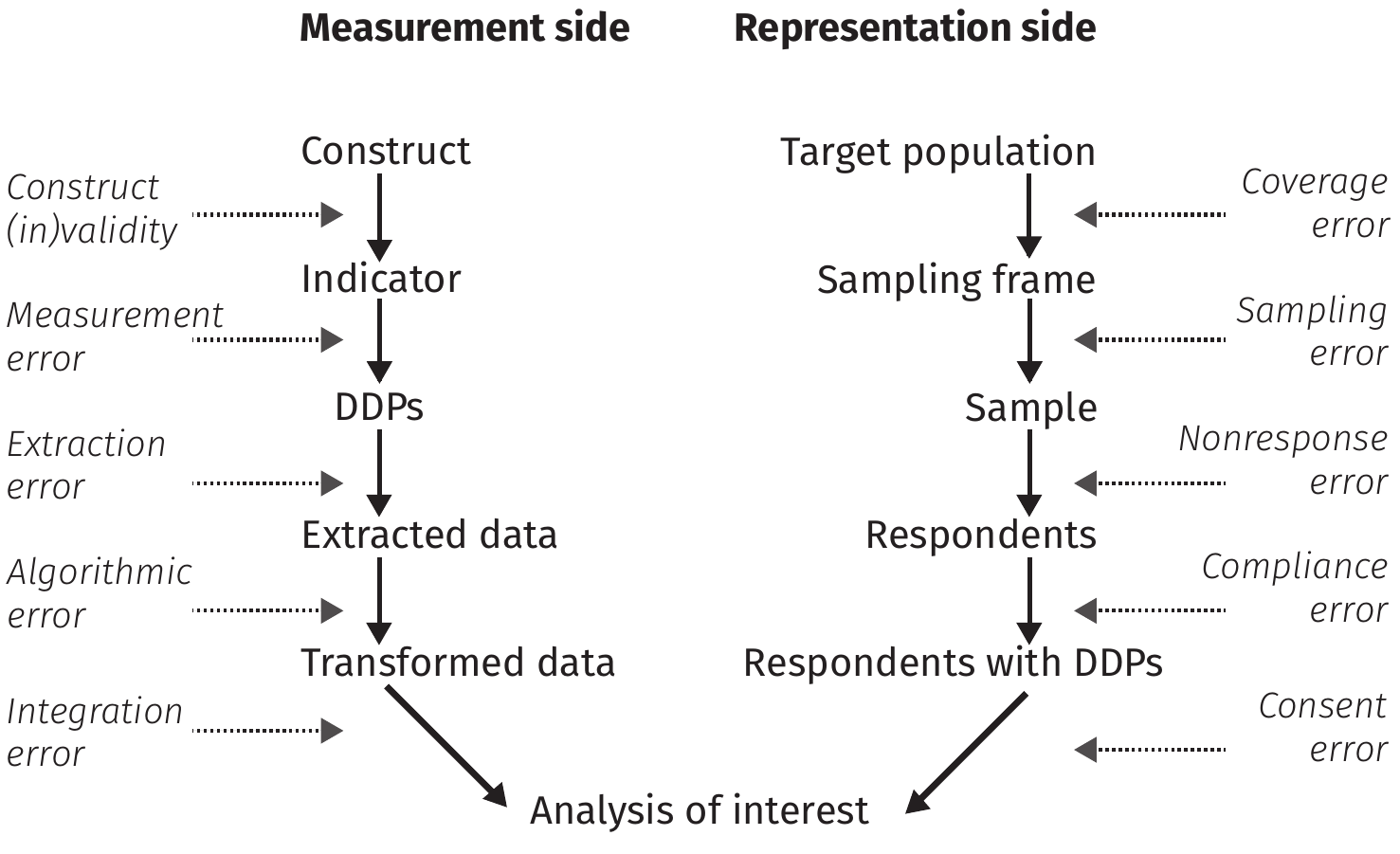}
    \caption{``Total error framework'' for social-scientific data collection with DDPs. Each step in the data collection process is shown, together with the  errors resulting from this step. Subsequent processing, modeling, and inference steps  \citep{amayaTotalErrorBig2020} are omitted.}
    \label{fig:DDP-total-error-general}
\end{figure}

\begin{figure}[tb]
    \includegraphics[width=\textwidth]{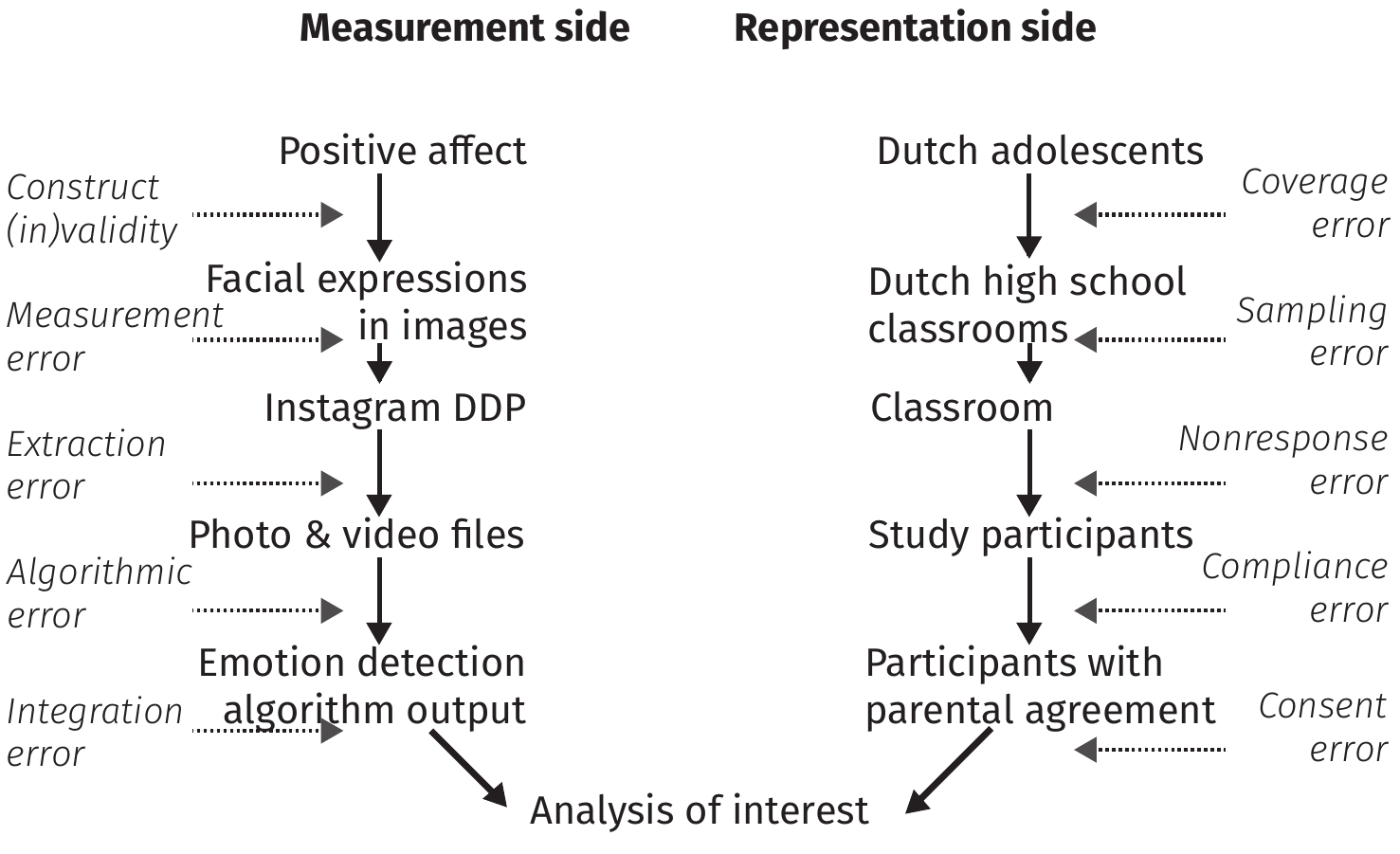}
    \caption{DDP total error framework applied to the example. Measurement side shows measurement of emotion (positive affect) from Instagram photos and videos. Note that the errors made on the measurement side of ``home'' vs. ``other'' location status are not shown here, although they will be present and affect the analysis of interest as well.}
    \label{fig:DDP-total-error-example}
\end{figure}

Here, we aim to aid future researchers in performing high-quality studies using DDPs by presenting a total error framework targeted specifically  at DDP collection. Figure \ref{fig:DDP-total-error-general} presents a generic overview of our framework. In addition, Figure \ref{fig:DDP-total-error-example} applies the framework from Figure \ref{fig:DDP-total-error-general} to our example research question above. As shown in Figures \ref{fig:DDP-total-error-general} and \ref{fig:DDP-total-error-example}, and following the standard TSE formulation, data collection consists of a ``measurement side'' and a ``representation side''. The measurement side deals with the extent to which the construct of theoretical interest is adequately measured by the procedure performed in the study. In a survey, this amounts to the extent to which answers to a survey question correspond to the construct of interest (e.g. well-being). With DDP collection, several additional steps are necessary, including definition of the construct, routine registration in the DDP, and extraction and transformation of the DDP into a variable to be analyzed. On the representation side, as with a standard survey, a population must defined, a sampling frame obtained, and respondents invited to participate. With DDP collection, additional steps are involved, which will lead to further respondent attrition. 

The following describes the steps of the framework in more detail. Throughout, we refer to Figure \ref{fig:DDP-total-error-general} and our hypothetical example illustrated (in part) by Figure \ref{fig:DDP-total-error-example}. 

\subsection{Measurement side}

\subsubsection{Construct}

On the measurement side of the framework, the first step is to consider how the \emph{constructs} (concepts) of interest can potentially be measured using \emph{indicators} (proxies) found in DDPs. Following our example, it would appear reasonable to presume that it is possible to determine whether a person is at home using location data, and indeed \citet{eleveltWhereYouUsing2019} showed that this can be done relatively reliably. Similarly, the existence of the field of ``affective computing'' suggests it may be possible to determine a person's emotions from their facial expressions in photos and videos \citep{dibekliogluRecognitionGenuineSmiles2015,kayaVideobasedEmotionRecognition2017,liDeepFacialExpression2020}. 

At this stage, errors can occur due to a mismatch between the chosen concept and the chosen indicator. For example, Instagram is often described as a ``storytelling'' device to assert the user's desired identity in contrast to the user's \emph{true} identity \citep[e.g.][]{martinez-garciaImagenEraDigital2017}. In other words, Instagram photos and videos are likely to measure how adolescents wish to be seen by others -- a construct that, as attested by popular culture, centuries of literature, and many readers' personal experience,  may differ from their genuine emotional state. 

\textbf{Construct error} is especially important since it enters at the very first step of measurement and has the potential to invalidate all downstream efforts unless controlled \citep{sarisDesignEvaluationAnalysis2007}. Methods of controlling construct error might include: careful elaboration of the theory underlying the research question, expert evaluation of the proposed indicator, and ``triangulation'' \citep{munafoRobustResearchNeeds2018} -- for example, comparison of research results between DDP and other types of measurement, or simultaneous DDP-survey measurement followed by multitrait-multimethod modeling \citep{oberski2017evaluating, revillaUsingPassiveData2017}. Because construct validity is such a crucial issue, we would suggest that simultaneous measurement using a combination of sources, including DDPs, is advisable; this idea is in line with similar advice given by \citet{japec2015big} and \citet{konitzerMeasuringNewsConsumption2020}. Our proposed workflow foresees in this need explicitly, by embedding the DDP collection step within a larger, more traditional, survey data collection effort. 

\subsubsection{Indicator}

Once the researcher has identified valid \emph{indicators} for the construct(s) of interest, the next step is to determine from which ``data controller(s)'' the \emph{DDP(s)} is/are most useful to answer the research question. For our example research question, we are interested in whether adolescents feel different emotions when they are at home compared to when they are not. As individuals typically switch locations multiple times a day, a DDP that registers location only once a day would not be sufficient to make the distinction we are interested in. The location history listed in the Instagram DDP only logs a location when it is selected by the user while sharing media on the ``timeline'' or in the ``stories'' \citep{manikonda2014analyzing} and would therefore not be sufficiently dense to appropriately distinguish between being home or not for every location the respondent visits throughout a day. Alternatively, Google Location History passively logs visited locations by combining internal phone GPS with connected WiFi devices and cell towers \citep{ruktanonchai2018using} and is therefore much more appropriate for the research question under evaluation. In terms of measuring emotions via social platforms \citep{kramer2014experimental}, adolescents frequently use Instagram \citep{valkenburg2011gender}, where emotions can be shared through both images and text \citep{bouko2020emotions}, which can be shared both publicly and privately.

At this stage, errors can occur when the measurements collected in the DDP diverge for some reason from what they intend to measure. For example, when satellites are temporarily out of order \citep{andrei2020signal}, the measurements logged in Google Location History might diverge more from the user's true location. 

\textbf{Measurement error} is particularly relevant because all measurements can be prone to error \citep{brakenhoff2018measurement} and it can distort all relationships under evaluation \citep{biemerIntroductionSurveyQuality2003}. A way to control for measurement error is by collecting multiple independent measurements of the construct of interest and investigate the variance of these measurements \citep{carroll2006measurement} or their correlations \citep{bland1996measurement}. Furthermore, these independent measurements can be used to estimate the unobserved ``true'' variable \citep{biemer2011latent}. In practice, this can be accounted for similarly as construct error, namely to supplement DDPs with survey measurements. In addition, measurement and construct error can be simultaneously estimated and accounted for using the previously discussed multitrait-multimethod modeling \citep{oberski2017evaluating}. To investigate positive affect using images in Instagram DDPs, a way to account for measurement error here can be to measure facial expression from other sources, such as self reports, sharing of selfies through ESM or using another DDP.  The information extracted from these different sources can then be used as indicators of the construct of interest by means of a latent variable model. 

\subsubsection{DDPs}

Once a specific set of \emph{DDPs} has been chosen to answer the research question of interest, the next step is to think more specifically which files of these DDPs are essential and how these relevant files are going to be \emph{extracted} from the DDPs. For our example research question, we are interested in determining the emotional expressions of faces on images. The extraction step here would be to identify all images in the Instagram DDP. 

\textbf{Extraction error} occurs when errors are made by the extraction algorithm, the image detection algorithm for example. A simple example of such an error is when the image detection algorithm only selects files with a .jpg extension, while the DDP of interest also contains images with a .png extension. Another example of this type of error is that images might be stored under different directories, and some are systematically missed. Moreover, controllers might also differentiate in their responses over time/location as well as file formats/structures\citep{ausloos2019getting}.
To minimize the possibility of extraction error, researchers should extensively investigate the content of the DDPs of interest, and how the structures and data type might differ over different users.

\subsubsection{Extracted data}

Once the relevant files have been \emph{extracted}, an algorithm can be applied \emph{transforming} the extracted files into data that can be used to answer the research question. In some cases, this step is very simple as data can be extracted from the files directly without further processing. For example, if the researcher is merely interested in seeing how often a person is home, this information is  directly available in Google DDPs, at the time of writing in the file "Location History.json". However, even with direct measures such as these, some degree of transformation might be needed; for example, as Google is not always certain of a person's location, multiple semantic locations are typically listed with corresponding probabilities. A predefined transformation rule can for example select the semantic location with the highest probability and the transformed data only contains these selected locations. A more complex set of transformations may also be needed, for example in the form of a pretrained supervised prediction model.  Following the example, a face detection algorithm \citep{hjelmaas2001face, hsu2002face} followed by an emotional expression detection algorithm could be applied to the images in the Instagram DDP, for example using pretrained models or by models further developed my means of transfer learning, such as by \citet{kaya2017video}.

\textbf{Algorithmic error} occurs when errors are made while generating transformed data from the extracted DDP files. When classifying emotions from faces, algorithmic error can be due to a face not being detected (as can be seen in Figure \ref{fig:illustration}, a face incorrectly being detected, an incorrect emotional classification, or because the algorithmic uncertainty is lost once a classification is made. In other words, algorithmic error is the typical classification or prediction error in predicting social variables using found data, which is the focus of a large body of literature \citep{kosinskiPrivateTraitsAttributes2013, blondelSurveyResultsMobile2015, jungherrAnalyzingPoliticalCommunication2015,settanniPredictingIndividualCharacteristics2018,eleveltWhereYouUsing2019}. In line with our exemplary research question, research has also illustrated that algorithmic error can influence outcomes of computer vision algorithms \citep{buolamwini2018gender}. In the current work, we emphasize that, while this type of error is certainly important, it constitutes only one type of error within the total error framework. In other words, the ``ground truth'' employed by supervised modeling exercises is, within our framework, an error-prone and potentially partially invalid proxy of the concept of interest.

Algorithmic error in the current framework is essentially prediction error on an (error-prone) measure of some socially relevant variable. As such, it is among the most studied errors within the framework at the time of writing. As emphasized in every basic textbook on machine learning, a proper evaluation of the likely amount of error is key, and can be accomplished by separating training and test observations, whether this is using data splits or resampling techniques \citep{bishop2006pattern,murphy2012machine,bengio2017deep}. When applying pretrained models as an extraction method, the researcher should ideally evaluate whether the error incurred within the DDP dataset at hand is indeed similar to that within the test set of the original model. For example, the type of photographs taken by teenagers might be different from standard benchmark datasets on which image recognition models were trained. Obtaining an accurate estimate of the algorithmic error rate also makes it possible to handle downstream decisions more adequately, by using standard measurement error models. For example, when we know that a classification model has a 90\% sensitivity and 75\% specificity, a simple table of predicted counts from this model can be corrected by multiplying it by the inverse of a matrix with these rates on the diagonal \citep{boeschoten2018updating,beauxis2017}. However, the difficulty of obtaining appropriate estimates of algorithmic errors should not be underestimated, particularly if pre-trained models are used for which training occurred using a different data-set.

\subsubsection{Transformed data} 

After the \emph{transformed data} files from all respondents are received and safely stored by the researcher, an integrated dataset can be generated containing data from all respondents, and linking the measurements received from possibly multiple DDPs to, for example, survey outcomes. As typically measurements at different time-points are collected through DDPs, attention should be paid to appropriately integrating the multiple datasets by linking on both person and time level \citep{harron2015methodological, zhang2012topics}. For our example research question, we should link the collected emotions to collected locations on time-level per person.

While linking the multiple sources on subject level and linking the subjects, \textbf{integration error} can occur \citep{kim2020data}, for example when time-stamps are not appropriately matched or when information collected from multiple sources is not appropriately linked on subject level \citep{doidge2019reflections}. Such errors can be prevented to an extent by creating software tests and other checks \citep{myers2004art} at every stage of the linkage process that create reports which can be compared with sensible expectations. For example, the time period should not suddenly extend into unseen years, outliers should be detected, etc. In addition, the procedures used should be computationally reproducible, so that any errors can be detected in the future and easily corrected \citep{stodden2014best,stodden2014implementing}.

\bigskip 

\begin{figure}[htp]
    \centering
    \includegraphics[width=\textwidth]{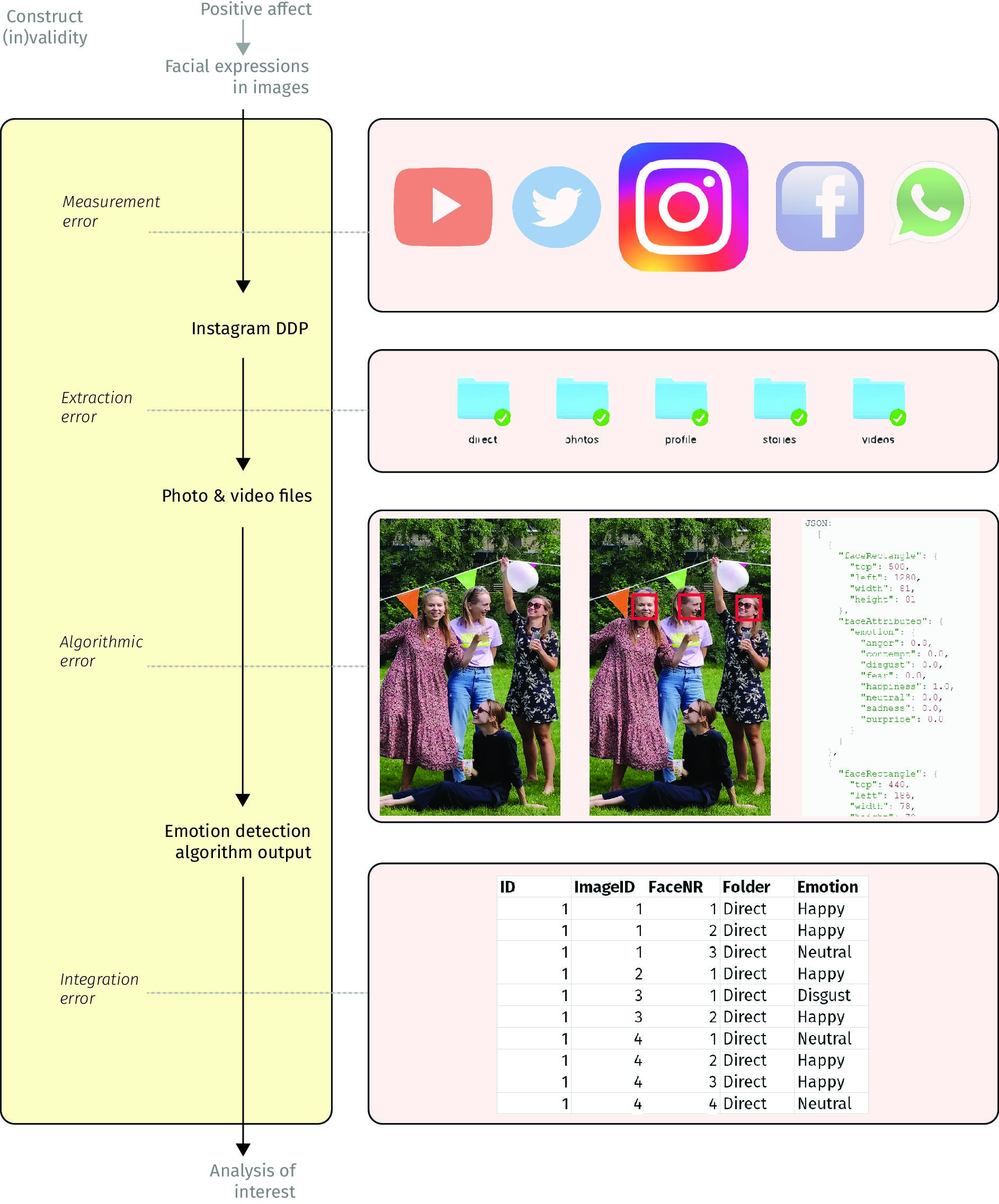}
    \caption{Visual illustration of the measurement side of the framework when using emotional expressions on images found in an Instagram DDP to measure positive affect.}
    \label{fig:illustration}
\end{figure}

See Figure \ref{fig:illustration} for a visual representation of how errors can affect outcomes on the measurement side of the framework. In addition, see the first part of Appendix \ref{app:checklist} for guidance on how severe bias due to measurement errors can be prevented.

\subsection{Representation side} 
\subsubsection{Target population}

On the representation side, researchers have in mind to what population their results should be generalized, a \emph{target population}. For the example research question, the target population is Dutch adolescents. Furthermore, researchers investigate how a sample or participants can be selected from that target population, this is the \emph{sampling frame}. If your target population is Dutch adolescents, it can be infeasible to randomly select a set of respondents out of that complete population directly. A practical approach can be to first select a sample of high schools and then select a number of adolescents here. Such a sampling scheme is known as clustered sampling \citep{lohr2008coverage,  bethlehem2011handbook}. 

The discrepancy between the target population and the sampling frame is denoted as \textbf{(under)coverage error}, as certain subgroups are not covered by the sampling frame. Coverage error can result in the problem that the obtained results cannot be generalized to the population of interest. For example, when Dutch high schools are used for the sampling frame, the subgroup of adolescents not going to high school have no probability of being included in the sample and obtained results can therefore not be generalized to Dutch adolescents, but only to Dutch adolescents going to high school. A solution can be to use multiple sampling frames \citep{lohr2009multiple}. 

\subsubsection{Sampling frame}

When the \emph{sampling frame} has been determined, the \emph{sample} can be selected using traditional sampling theory \citep{cochran2007sampling}, for example a simple random sample can be selected by randomly selecting a number of adolescents from the high school registers and invite them to participate in the research. Alternatively, using strata or clusters can be more convenient here, for example to first select a number of high schools and approach a sample of adolescents via these schools. 

Failing to select a representative sample results in \textbf{sampling error}, failure to generalize results to the target population. Many large studies use model-based approaches \citep{chambers2003analysis}, combining several stages of stratification and clustering to minimize sampling error \citep{de2008international}. Alternatively, adaptive designs can be used to minimize sampling error \citep{bethlehem2011handbook} and to for example increase the sampling probability for certain subgroups if their response rate is relatively low. Stratification has also been listed by \citet{japec2015big} as an important contribution to the goal of generalizability in big data research. 

\subsubsection{Sample}

Once the \emph{sample} has been determined, its members can be invited to participate in the research. As with any type of research, part of the sampled subjects will not or only partly respond. This can be due to multiple reasons. First, the subject is not willing to participate at all. Second, the subject is willing to participate in the overall project, but is not willing to provide her DDP. Third, the subject is willing to participate, but does not use the platform from which the DDP is requested.

Regardless of the reason for not participating in the research, this will lead to \textbf{nonresponse error} an can lead to bias in results \citep{groves2008impact}. To minimize bias caused by respondents not willing to participate or only willing to partly participate, it is recommended to accompany data-download research with questionnaires. This provides the researcher with substantive information regarding the non-responders in terms of data-download packages. When viewed as a missing data problem, this means that once more information is known about the non-respondents, the likelihood increases that the Missingness is At Random (MAR), as variables are observed through which the missingness can be explained. This is in contrast to a situation when nothing is known about the nonrespondents, so that the missingness cannot be explained (Missing Not At Random, MNAR) \citep{schafer2002missing}. When the missingness can be explained, it can be accounted for by method such as multiple imputation or weighting \citep{boeschoten2017obtain}. To minimize the number of respondents that are willing to participate but to not use the platform under investigation, the researcher should also focus on how often the target population makes use of the platform under investigation when determining which platform to use for research. When considering the example research question, existing research showed that YouTube, WhatsApp, Instagram and Snapchat were used most frequently by adolescents in 2019 \citep{van2019urban}. Furthermore, Android had a market share of 86.1\% in 2017 \citep{ahvanooey2020survey}.

\subsubsection{Respondents}

If a respondent decides to participate in the research, she still needs to work through a process of multiple stages. The packages should be requested and downloaded. A piece of software should be installed and the packages should be opened and processed with this software, as can be seen in Figure \ref{fig:high-level}. Next, the output is generated by the software and the respondent determines whether she is willing to share this output with the researcher and, if so, actually approve the sharing. 

These steps are not straightforward. Therefore, clear guidelines, reminders and assistance are required to guide the respondents through this process \citep{shirima2007use}. Some attrition is likely to occur due to the fact that respondents are not willing or  able to invest the time and effort in this procedure, resulting in \textbf{compliance error}. For example, when a respondent requests her Instagram DDP, it typically takes several hours to days for Instagram to prepare this DDP, so the respondent needs to reserve multiple moments throughout several days to successfully participate in this research, and the researchers should probably build in several reminders throughout this process to nudge the respondent into successfully completing the process. Furthermore, by processing and visualizing locally, the respondent has control over the data and is truly informed.

\subsubsection{Respondents with DDPs} 

Once the respondent complied with all the steps required to complete the process, the transformed data is collected in a file. For our example research question, a respondent will for example review a csv file containing timestamps, classified emotions and supplementary information describing whether it was text or an image that was classified (as can be seen at the bottom of Figure \ref{fig:illustration}.

This file should be reviewed by the respondent in order to give informed consent regarding sharing this information with the researchers. If the respondent decides to not or only partly share this file, this results in \textbf{consent error}. Consent error may be substantial, and could be related to topics of interest measured within the DDPs. Without any further information about the respondent, for example for a survey, this would lead to missingness ``not at random'' \citep[MNAR;][]{schafer2002missing}, which is difficult to account for. With  information from surveys or other sources, it may be reasonable to assume the missingness is ``at random'' (MAR), especially when survey variables are strongly related to the study outcomes. 

``Local signal processing'' may alleviate  the consent error considerably. First, local processing will allow researchers to avoid requesting sensitive information, perhaps making respondents more willing to share \citep{singer1993informed}. For example, respondents could be more likely to give consent to share the datum of ``looking unhappy'' in a photograph than sharing all their private images. Second, the respondent can see that the only  information that is requested is directly related to their interaction with the researcher: a scientific study. Most adolescents will intuit that, to study well-being, the researcher does not need to know their study habits, for instance. In other words, local signal processing is designed to comply with key data protection principles such as `data minimization' and `data protection by design' as well as more generally preserve the interaction's ``contextual integrity" \citep{nissenbaum2004privacy}. Some studies have suggested that preserving  contextual integrity can help improve consent \citep{Hutton2015IDS}.

Once the integrated data-set is finalized, it can be used to perform the final analyses  to answer the research question of interest. For example, the researcher can investigate what type of emotions are more often detected while being at home and while being at other locations, and it can be investigated how these differences in emotional outings differ within and between persons.

\bigskip 

See the second part of Appendix \ref{app:checklist} for guidance on how severe bias due to representation errors can be prevented.

\section{Discussion}

Data-download packages (DDPs) allow us to study known phenomena  in a novel manner, or even to study new social phenomena. Using DDPs  for scientific research is attractive for multiple reasons. First, the existence of DDPs, and the right of the data subject to pass on information to social scientists, is guaranteed by EU law. Second, participants can easily investigate the data they share to give informed consent. Third, by starting off with a traditional random sample, the approach suggested in this paper allows researchers to generalize to populations of interest more easily than could be achieved with ``found samples''. This approach also allows for longitudinal data collection in parallel with the DDPs. More generally, fourth, DDPs do not only provide a very diverse set of available digital traces, but they can also easily be combined with other data, such as other DDPs, surveys, register data, and so forth. 
Finally, the approach suggested in this paper allows for experimental designs using digital trace outcomes, but under the same scrutiny as regular social-scientific experiments and with true informed consent that respects the contextual integrity of the research-participant interaction. 

Of course, use of DDPs is also challenging. We have focused on summarizing some of the challenges to inference within our error framework, and hope this framework can serve as a guide to preventing errors where possible, and mitigating their effects otherwise. At the same time, the suggested approach also has several drawbacks that are unrelated to inference per se. 

First, researchers should have good faith in not only respondents, but also in data controllers, as both have the opportunity to omit data during the process. Data controllers can fail to provide respondents their complete DDPs, and respondents can choose to remove parts of the DDP they are not willing to share with the researcher. As data not shared with the researcher can differ from shared data, this can bias results. A second challenge is that the world of DDPs changes rapidly. The structure and content changes continuously and individuals can be triggered to delete their own packages making them useless as research subjects. A third challenge is that, to safeguard participants’ privacy and for scientists to comply with data protection requirements themselves, most research infrastructure should be set up in advance. For example, it should be clear which parts of which data-download packages are selected and an algorithm should be prepared to make transformations to a pre-defined format. A fourth disadvantage is that available and free pre-trained algorithms are not always available for the specific research purposes, requiring the researcher to collect raw data and train an algorithm. Fifth, digital skills of participants are a major challenge. To address them an easy to use front end of the data collection tool is key. A sixth challenge is that data-download packages are not consistently formatted over different data controllers. For example, there are already many ways to provide timestamps \citep{dyreson1993timestamp} so software should be adjusted to appropriately handle such differences. A seventh challenge is that a DDP itself is not formatted as a typical data set with respondents as rows and variables as columns. Instead, it typically comes as a zip file containing json files, images and videos, and processing should take place in order for it to be used for statistical analyses. A last challenge is that conducting research of this type should be carried out by a multidisciplinary team of social scientists, data scientists, computer scientists and data management experts.

Researchers can minimize the influence of issues such as the rapidly changing environment of DDPs and the inconsistency in DDPs by focussing their processes on structural characteristics such as for example usernames and timestamps. Issues such as setting up the infrastructure in advance and training algorithms without access to the complete data have been overcome before \citep{lovestone2020european}, however ensuring that the usability of such infrastructures meets the level of digital skills of the participant remains an important challenge here. For challenges regarding data protection, informed consent, reproducibility and replicability, extensive research has been performed and guidelines have been developed on which we reflect in the following subsections.

\subsection{Data protection and informed consent}

Before a DDP of a respondent is shared, it is unknown what kind of information the package exactly contains. Social researchers will  only be interested in the specific parts of the DDP that help to answer their research question, but a DDP possibly contains sensitive personal information. By using distributed local computation at the respondent's device to extract only the relevant information, it can be prevented that a researcher stores sensitive information. For example, an Instagram DDP can contain sensitive images. The researcher is not interested in the sensitive content, but in the emotional expressions of the faces on these images. Therefore, an emotional detection algorithm could be run locally and only the classifications of the emotional expressions are shared with the researcher.

During this privacy preserving transformation step, three aspects should be carefully considered. First, respondents store their DDPs locally on a device. After participating, respondents should be informed of this and should have the option to either preserve the packages under their own responsibility, or to permanently delete the packages from the device in use. 
Second, to maximize informed consent, respondents should be shown an example illustrating what information is extracted from the data-download package. In the case of transforming faces on pictures into classifications of emotional expressions, the example should show a picture as input and the classifications of emotional expressions per detected face at output, as can be seen in Figure \ref{fig:illustration}. Such an example makes clear what information from the data-download package is shared with the researcher exactly. In addition, respondents should have access to output of the transformations applied to their own DDP to explicitly approve or reject sharing the transformations with the researcher. Existing research on successful informed consent can be consulted, see for example \citep{kreuter2016framing}.

To ensure that sensitive information is not shared with the researcher and to ensure that the procedure of obtaining the transformed data occurs in a privacy preserving and ethical way, it is important that researchers consult ethical review boards of their universities in this process and obtain ethical approval for the research. Furthermore, researchers should consult data managers to develop a solid plan to receive the transformed data in a safe environment from the respondents and to generate an integrated database built with an architecture that can be accessed by the researchers, such as SURFsara in the Netherlands \citep{scheermansecure2020}.

\subsection{Reproducibility and replicability}
Although reproducibility and replicability are essential for scientific research \citep{patil2016statistical,stodden2014best}, these criteria are challenging to meet when using DDPs \citep{gayo2012no}. The field of digital trace data in general is a rapidly changing environment \citep{stier2019integrating}, and this holds for DDPs as well. When using local computation, reproducibility may only be feasible on the level of the transformed data received by the researchers, not on the raw DDPs, as they were never in the possession of the researcher in the first place. 

To support replicability, tools and analysis code should depend on structures specific for particular data controllers as little as possible, and should be easily updatable and extendable as structures of DDPs from specific data controllers will inevitably change. To help achieve this goal, the highest standards of software engineering for architecture design, testing, documentation, version control and support should be applied and software engineers should be involved during all stages of the process \citep{myers2004art}. In addition, FAIR principles \citep{wilkinson2016fair} should be used for data archiving, documentation and long-term storage. As these go beyond the expertise of most social scientists, Research Data Management Offices should be involved or at least consulted, see for example \citep{UURDMS}. Frameworks such as differential privacy \citep{dwork2008differential} are relevant to guarantee reuse. 

\section{Conclusion}

If researchers interested in using DDPs for scientific research follow the proposed workflow, improvements can be made regarding generalizability of findings. This holds for the exemplary research question discussed, but also for example for the research questions discussed in the introduction such as the network analysis from mobile phone data \citep{blondelSurveyResultsMobile2015}, price indexing from online shops \citep{de2013online}, political opinion and electoral success prediction from Twitter data \citep{schoenPowerPredictionSocial2013, jungherrAnalyzingPoliticalCommunication2015}, and personality profiling from Facebook ``likes" (\cite{kosinskiPrivateTraitsAttributes2013}. Furthermore, research questions typically investigated using surveys can be executed without suffering from issues such as recall bias or bias due to social desirability, such as the examples discussed in the introduction regarding such as energy consumption \citep{guerra2010occupants}
time spent \citep{eleveltWhereYouUsing2019} or budget research \citep{breedveld2002background}.

To summarize, it is clear that our proposal is no silver bullet for solving all problems associated with modern social science. In spite of these challenges, however, we believe that leveraging the advantages of DDP collection can become an important tool in the social scientist's arsenal. 
\newpage
\appendix 

\section{Checklist for social scientific research using Data Download Packages}
\label{app:checklist}

\subsection*{Measurement side}
(determine per construct separately)

\paragraph{Construct}
\begin{itemize}
    \item[$\square$] The construct of interest is clearly defined 
    \item[$\square$] The construct of interest matches the scope of the research
\end{itemize}

\paragraph{Indicator(s)}
\begin{itemize}
    \item[$\square$] All aspects of the construct can be sufficiently represented through observable indicators (proxies)
    \item[$\square$] The indicators can be measured by data controllers
\end{itemize}

\paragraph{DDPs}
\begin{itemize}
    \item[$\square$] Data controllers are selected in which the indicators of interest are measured
    \item[$\square$] The denseness of the measured indicators matches the research purpose
    \item[$\square$] The credibility of the data controller is positively evaluated
    \item[$\square$] The number of different data controllers is minimized to reduce response burden
\end{itemize}

\paragraph{Extracted data}
\begin{itemize}
\item[$\square$] Presence of the indicator is evaluated for all file formats present in the DDP
\item[$\square$] Relevant files are extracted using validated scripts with known accuracy rates
\end{itemize}

\paragraph{Transformed data}
\begin{itemize}
    \item[$\square$] A transformation method is selected that extracts the outcome values for each indicator
    \item[$\square$] The transformation method is trained on a sample similar to the data collected by means of DDPs
    \item[$\square$] The transformation method has a known accuracy rate estimated on a comparable data-set
    \item[$\square$] The transformation method does not systematically include, exclude or misclassifies specific (identifiable) cases
    \item[$\square$] The outcome values sufficiently represent all indicators identified
\end{itemize}

\paragraph{Analysis of interest}
\begin{itemize}
    \item[$\square$] The shared data is linked on person level, such that different sets of transformed data are represented by different columns in one data-set
    \item[$\square$] Individual respondents can be clearly identified, for example by means of an anonymized identification number 
   \item[$\square$] The variables are clearly identified for each respondent
\end{itemize}

\subsection*{Representation side}
\paragraph{Target population}
\begin{itemize}
    \item[$\square$] A target population is identified that matches the research purpose
    \item[$\square$] All identifiable subgroups can in theory be included in the study
\end{itemize}

\paragraph{Sampling frame}
\begin{itemize}
    \item[$\square$] All identifiable subgroups of the target population are present in the sampling frame
    \item[$\square$] Evaluate whether the available sampling frame matches the research purpose
\end{itemize}

\paragraph{Sample}
\begin{itemize}
    \item[$\square$] All subgroups in the sampling frame have a probability to be included in the sample
    \item[$\square$] All subgroups in the sampling frame have an equal or known probability to be included in the sample 
\end{itemize}

\paragraph{Respondents}
\begin{itemize}
    \item[$\square$] The communication towards the sample is clear and simple
    \item[$\square$] Communication is possible in the respondent's language
    \item[$\square$] The procedure is explained in a step-by-step manner for informed consent at the start of the procedure
\end{itemize}

\paragraph{Respondent's DDPs}
\begin{itemize}
    \item[$\square$] The software's usability has been validated on an independent validation sample
    \item[$\square$] The software is available for different types of devices and different versions of operating systems
    \item[$\square$] 24 hour assistance is available during the data collection period
\end{itemize}

\paragraph{Analysis of interest}
\begin{itemize}
    \item[$\square$] The respondents can see the final data-set containing the transformed data before it is shared with the researcher for informed consent
\end{itemize}

\section{Definitions}
\begin{itemize}
\item Personal data: Information relating to an identified or identifiable natural person \citep{van2020general}
Data subject: The person that the personal data refer to 
Data processing entity / Data controller: The person or organization responsible for processing personal data. In this paper we refer to the online platforms providing data download packages as data controllers. However, note that as a researcher collecting DDPs, you are a data controller as well \citep{van2020general}

\item Data controller: The person or organization responsible for processing personal data. The controller decides which data will be processed, how and why \citep{van2020general}.

\item Data download package (DDP): Because of the right of data access, data subjects are always allowed to retrieve their personal data from data controllers. Here, data controllers are obliged to comply with such a request and because of the right of data portability, provide the requested data in a machine readable format. To comply with these rules, social media platforms typically provide data subjects with a .zip file containing the personal data requested \citep{van2020general}

\item Consent: When data subject provide researchers their DDPs, consent should be provided. This means that the data subject confirms that the data provided given freely; that the data subject is informed regarding what data are shared exactly and how the data will be processed by the researcher. Consent can be provided via a written, electronic or oral statement \citep{van2020general}

\bigskip

\item Target population: The population to be investigated, and about which conclusions are to be drawn \citep{bethlehem2011handbook}
\item Sampling frame: A list, map, or other specification of units in the target population from which a sample of data subjects may be selected \citep{de2008international}
\item Sample: The set of data subjects within the sampling frame selected for participation in the research in practice
\item Respondents: Data subjects within the sample who complied with participation in the research
\item Responses: Data collected from the data subjects who complied with participation in the research

\bigskip

\item Construct: A conceptual variable that is known to exist but cannot be directly observed \citep{privitera2018research}

\item Indicator: Variables (constructed by means of measurement instruments) that aim to measure either the construct of interest or are closely related to the construct of interest

\item Transformation method: Algorithm that is used to transform the data obtained from the DDPs into features and classifications that can be used for further research

\item Transformed data: The features or classification extracted using transformation method which can be used for further research

\item Data integration: The theory and techniques used for data linkage and micro integration. Here, data linkage techniques vary from record linkage to statistical matching. Micro integration techniques vary from harmonization of measures in concept to actual adjustments of data \citep{zhang2012topics}.

\end{itemize}

\newpage

\printbibliography

@book{bishop2006pattern,
  title={Pattern recognition and machine learning},
  author={Bishop, Christopher M},
  year={2006},
  publisher={springer}
}

@book{murphy2012machine,
  title={Machine learning: a probabilistic perspective},
  author={Murphy, Kevin P},
  year={2012},
  address={MA},
  publisher={MIT press}
}

@book{bengio2017deep,
  title={Deep learning},
  author={Bengio, Yoshua and Goodfellow, Ian and Courville, Aaron},
  year={2017},
  address={MA},
  publisher={MIT press}
}

@INPROCEEDINGS{beauxis2017,  
author={E. {Beauxis-Aussalet} and L. {Hardman}},  
booktitle={2017 IEEE International Conference on Data Science and Advanced Analytics (DSAA)},   
title={Extended Methods to Handle Classification Biases},   year={2017},  
volume={},  
number={},  
pages={765-774}
}

@book{myers2004art,
  title={The art of software testing},
  author={Myers, Glenford J and Badgett, Tom and Thomas, Todd M and Sandler, Corey},
  volume={2},
  year={2004},
  publisher={Wiley Online Library}
}

@article{stodden2014best,
  title={Best Practices for Computational Science: Software Infrastructure and Environments for Reproducible and Extensible Research},
  author={Stodden, Victoria and Miguez, Sheila},
  journal={Journal of Open Research Software},
  volume={2},
  number={1},
  year={2014},
  publisher={Ubiquity Press}
}

@inproceedings{Hutton2015IDS,
  title={``I Didn't Sign Up for This!": Informed Consent in Social Network Research},
  author={Luke Hutton and T. Henderson},
  booktitle={ICWSM},
  year={2015}
}

@article{nissenbaum2004privacy,
  title={Privacy as contextual integrity},
  author={Nissenbaum, Helen},
  journal={Wash. L. Rev.},
  volume={79},
  pages={119},
  year={2004},
  publisher={HeinOnline}
}

@book{stodden2014implementing,
  title={Implementing reproducible research},
  author={Stodden, Victoria and Leisch, Friedrich and Peng, Roger D},
  year={2014},
  publisher={CRC Press}
}

@article{konitzerMeasuringNewsConsumption2020,
	title = {Measuring {News} {Consumption} {With} {Behavioral} {Versus} {Survey} {Data}},
	issn = {1556-5068},
	url = {https://www.ssrn.com/abstract=3548690},
	doi = {10.2139/ssrn.3548690},
	language = {en},
	urldate = {2020-08-12},
	journal = {SSRN Electronic Journal},
	author = {Konitzer, Tobias and Allen, Jennifer and Eckman, Stephanie and Howland, Baird and Mobius, Markus M. and Rothschild, David M. and Watts, Duncan},
	year = {2020}
}

@article{revillaUsingPassiveData2017,
	title = {Using {Passive} {Data} {From} a {Meter} to {Complement} {Survey} {Data} in {Order} to {Study} {Online} {Behavior}},
	volume = {35},
	issn = {0894-4393, 1552-8286},
	url = {http://journals.sagepub.com/doi/10.1177/0894439316638457},
	doi = {10.1177/0894439316638457},
	language = {en},
	number = {4},
	urldate = {2020-08-12},
	journal = {Social Science Computer Review},
	author = {Revilla, Melanie and Ochoa, Carlos and Loewe, Germán},
	month = aug,
	year = {2017},
	pages = {521--536}
}

@article{munafoRobustResearchNeeds2018,
	title = {Robust research needs many lines of evidence},
	volume = {553},
	issn = {0028-0836, 1476-4687},
	url = {http://www.nature.com/articles/d41586-018-01023-3},
	doi = {10.1038/d41586-018-01023-3},
	language = {en},
	number = {7689},
	urldate = {2020-08-12},
	journal = {Nature},
	author = {Munafò, Marcus R. and Davey Smith, George},
	month = jan,
	year = {2018},
	pages = {399--401}
}

@book{sarisDesignEvaluationAnalysis2007,
	address = {Hoboken, N.J},
	series = {Wiley series in survey methodology},
	title = {Design, evaluation, and analysis of questionnaires for survey research},
	isbn = {978-0-470-11495-7},
	publisher = {Wiley-Interscience},
	author = {Saris, Willem E. and Gallhofer, Irmtraud N.},
	year = {2007},
	note = {OCLC: ocm80358503}
}

@book{martinez-garciaImagenEraDigital2017,
	title = {La imagen en la era digital},
	copyright = {Attribution-NonCommercial-NoDerivatives 4.0 Internacional},
	isbn = {978-84-17270-12-4},
	url = {https://idus.us.es/handle/11441/91571},
	language = {spa},
	urldate = {2020-08-12},
	publisher = {Egregius},
  address={Universidad de Sevilla},
	author = {Martínez-García, Ángeles},
	year = {2017}
}

@article{eleveltWhereYouUsing2019,
	title = {Where {You} at? {Using} {GPS} {Locations} in an {Electronic} {Time} {Use} {Diary} {Study} to {Derive} {Functional} {Locations}},
	issn = {0894-4393, 1552-8286},
	shorttitle = {Where {You} at?},
	url = {http://journals.sagepub.com/doi/10.1177/0894439319877872},
	doi = {10.1177/0894439319877872},
	language = {en},
	urldate = {2020-08-12},
	journal = {Social Science Computer Review},
	author = {Elevelt, A. and Bernasco, W. and Lugtig, P. and Ruiter, S. and Toepoel, V.},
	year = {2019}
}

@article{schoenPowerPredictionSocial2013,
	title = {The power of prediction with social media},
	volume = {23},
	issn = {1066-2243},
	url = {https://www.emerald.com/insight/content/doi/10.1108/IntR-06-2013-0115/full/html},
	doi = {10.1108/IntR-06-2013-0115},
	language = {en},
	number = {5},
	urldate = {2020-08-12},
	journal = {Internet Research},
	author = {Schoen, Harald and Gayo-Avello, Daniel and Takis Metaxas, Panagiotis and Mustafaraj, Eni and Strohmaier, Markus and Gloor, Peter},
	editor = {Gayo-Avello, Panagiotis Takis Metax, Daniel},
	month = oct,
	year = {2013},
	pages = {528--543}
}

@article{biemerTotalSurveyError2010,
	title = {Total {Survey} {Error}: {Design}, {Implementation}, and {Evaluation}},
	volume = {74},
	issn = {0033-362X, 1537-5331},
	shorttitle = {Total {Survey} {Error}},
	url = {https://academic.oup.com/poq/article-lookup/doi/10.1093/poq/nfq058},
	doi = {10.1093/poq/nfq058},
	language = {en},
	number = {5},
	urldate = {2020-08-12},
	journal = {Public Opinion Quarterly},
	author = {Biemer, Paul P.},
	month = jan,
	year = {2010},
	pages = {817--848}
}

@book{biemerIntroductionSurveyQuality2003,
	address = {Hoboken, NJ},
	series = {Wiley series in survey methodology},
	title = {Introduction to survey quality},
	isbn = {978-0-471-19375-3},
	publisher = {Wiley},
	author = {Biemer, Paul P.  and Lyberg, Lars},
	year = {2003},
	note = {}
}

@article{grovesTotalSurveyError2010,
	title = {Total {Survey} {Error}: {Past}, {Present}, and {Future}},
	volume = {74},
	issn = {0033-362X, 1537-5331},
	shorttitle = {Total {Survey} {Error}},
	url = {https://academic.oup.com/poq/article-lookup/doi/10.1093/poq/nfq065},
	doi = {10.1093/poq/nfq065},
	language = {en},
	number = {5},
	urldate = {2020-08-07},
	journal = {Public Opinion Quarterly},
	author = {Groves, R. M. and Lyberg, L.},
	month = jan,
	year = {2010},
	pages = {849--879}
}

@article{kayaVideobasedEmotionRecognition2017,
	title = {Video-based emotion recognition in the wild using deep transfer learning and score fusion},
	volume = {65},
	issn = {02628856},
	url = {https://linkinghub.elsevier.com/retrieve/pii/S0262885617300367},
	doi = {10.1016/j.imavis.2017.01.012},
	language = {en},
	urldate = {2020-08-12},
	journal = {Image and Vision Computing},
	author = {Kaya, Heysem and Gürpınar, Furkan and Salah, Albert Ali},
	month = sep,
	year = {2017},
	pages = {66--75}
}

@article{dibekliogluRecognitionGenuineSmiles2015,
	title = {Recognition of {Genuine} {Smiles}},
	volume = {17},
	issn = {1941-0077},
	doi = {10.1109/TMM.2015.2394777},
	number = {3},
	journal = {{IEEE} Transactions on Multimedia},
	author = {Dibeklioğlu, Hamdi and Salah, Albert Ali and Gevers, Theo},
	month = mar,
	year = {2015},
	pages = {279--294}
}

@article{liDeepFacialExpression2020,
	title = {Deep {Facial} {Expression} {Recognition}: {A} {Survey}},
	issn = {1949-3045, 2371-9850},
	shorttitle = {Deep {Facial} {Expression} {Recognition}},
	url = {https://ieeexplore.ieee.org/document/9039580/},
	doi = {10.1109/TAFFC.2020.2981446},
	urldate = {2020-08-12},
	journal = {{IEEE} Transactions on Affective Computing},
	author = {Li, Shan and Deng, Weihong},
	year = {2020},
	pages = {1--1}
}

@article{szellMultirelationalOrganizationLargescale2010,
	title = {Multirelational organization of large-scale social networks in an online world},
	copyright = {©  . Freely available online through the PNAS open access option.},
	issn = {0027-8424, 1091-6490},
	url = {https://www.pnas.org/content/early/2010/07/13/1004008107},
	doi = {10.1073/pnas.1004008107},
	language = {en},
	urldate = {2020-08-05},
	journal = {Proceedings of the National Academy of Sciences},
	author = {Szell, Michael and Lambiotte, Renaud and Thurner, Stefan},
	month = jul,
	year = {2010},
	pmid = {20643965},
	note = {Publisher: National Academy of Sciences
Section: Social Sciences}
}

@book{valliantPracticalToolsDesigning2018,
	address = {Cham},
	title = {Practical {Tools} for {Designing} and {Weighting} {Survey} {Samples}},
	url = {https://link.springer.com/10.1007/978-3-319-93632-1},
	language = {English},
	urldate = {2020-08-05},
	publisher = {Springer International Publishing : Imprint: Springer},
	author = {Valliant, Richard and Dever, Jill A and Kreuter, Frauke},
	year = {2018}
}

@techreport{ausloosGDPRTransparencyResearch2019,
	address = {Rochester, NY},
	type = {{SSRN} {Scholarly} {Paper}},
	title = {{GDPR} {Transparency} as a {Research} {Method}},
	url = {https://papers.ssrn.com/abstract=3465680},
	language = {en},
	number = {ID 3465680},
	urldate = {2020-08-05},
	institution = {Social Science Research Network},
	author = {Ausloos, Jef},
	month = may,
	year = {2019},
	doi = {10.2139/ssrn.3465680}
}

@article{settanniPredictingIndividualCharacteristics2018,
	title = {Predicting {Individual} {Characteristics} from {Digital} {Traces} on {Social} {Media}: {A} {Meta}-{Analysis}},
	volume = {21},
	issn = {2152-2715, 2152-2723},
	shorttitle = {Predicting {Individual} {Characteristics} from {Digital} {Traces} on {Social} {Media}},
	url = {http://www.liebertpub.com/doi/10.1089/cyber.2017.0384},
	doi = {10.1089/cyber.2017.0384},
	language = {en},
	number = {4},
	urldate = {2020-08-05},
	journal = {Cyberpsychology, Behavior, and Social Networking},
	author = {Settanni, Michele and Azucar, Danny and Marengo, Davide},
	month = apr,
	year = {2018},
	pages = {217--228}
}

@article{kosinskiPrivateTraitsAttributes2013,
	title = {Private traits and attributes are predictable from digital records of human behavior},
	volume = {110},
	issn = {0027-8424, 1091-6490},
	url = {http://www.pnas.org/cgi/doi/10.1073/pnas.1218772110},
	doi = {10.1073/pnas.1218772110},
	language = {en},
	number = {15},
	urldate = {2020-08-04},
	journal = {Proceedings of the National Academy of Sciences},
	author = {Kosinski, M. and Stillwell, D. and Graepel, T.},
	month = apr,
	year = {2013},
	pages = {5802--5805}
}

@article{kingNewModelIndustry2019,
	title = {A {New} {Model} for {Industry}–{Academic} {Partnerships}},
	issn = {1049-0965, 1537-5935},
	url = {https://www.cambridge.org/core/product/identifier/S1049096519001021/type/journal_article},
	doi = {10.1017/S1049096519001021},
	language = {en},
	urldate = {2020-08-05},
	journal = {PS: Political Science \& Politics},
	author = {King, Gary and Persily, Nathaniel},
	month = aug,
	year = {2019},
	pages = {1--7}
}

@article{Oberski2020Differential,

    journal = {Harvard Data Science Review},
    doi = {10.1162/99608f92.63a22079},
    number = {1},
    title = {Differential Privacy and Social Science: An Urgent Puzzle},
    url = {https://hdsr.mitpress.mit.edu/pub/g9o4z8au},
    volume = {2},
    author = {Oberski, Daniel L. and Kreuter, Frauke},
    date = {2020-01-31},
    year = {2020},
    month = {1},
    day = {31},

}

@article{kingEnsuringDataRichFuture2011,
	title = {Ensuring the {Data}-{Rich} {Future} of the {Social} {Sciences}},
	volume = {331},
	issn = {0036-8075, 1095-9203},
	url = {https://www.sciencemag.org/lookup/doi/10.1126/science.1197872},
	doi = {10.1126/science.1197872},
	language = {en},
	number = {6018},
	urldate = {2020-08-05},
	journal = {Science},
	author = {King, G.},
	month = feb,
	year = {2011},
	pages = {719--721}
}

@article{mellonTwitterFacebookAre2017,
	title = {{Twitter} and {Facebook} are not representative of the general population: {Political} attitudes and demographics of {British} social media users},
	volume = {4},
	issn = {2053-1680, 2053-1680},
	shorttitle = {{Twitter} and {Facebook} are not representative of the general population},
	url = {http://journals.sagepub.com/doi/10.1177/2053168017720008},
	doi = {10.1177/2053168017720008},
	language = {en},
	number = {3},
	urldate = {2020-08-05},
	journal = {Research \& Politics},
	author = {Mellon, Jonathan and Prosser, Christopher},
	month = jul,
	year = {2017},
	pages = {205316801772000}
}

@misc{messingFacebookPrivacyProtectedFull2020,
	title = {Facebook {Privacy}-{Protected} {Full} {URLs} {Data} {Set}},
	url = {https://doi.org/10.7910/DVN/TDOAPG},
	urldate = {2020-08-05},
	publisher = {Harvard Dataverse},
	author = {Messing, Solomon and DeGregorio, Christina and Hillenbrand, Bennett and King, Gary and Mahanti, Saurav and Mukerjee, Zagreb and Nayak, Chaya and Persily, Nate and State, Bogdan and Wilkins, Arjun},
	collaborator = {Mukerjee, Zagreb},
	year = {2020},
	doi = {10.7910/DVN/TDOAPG},
	note = {type: dataset}
}

@article{dorazioDifferentialPrivacySocial2015,
	title = {Differential {Privacy} for {Social} {Science} {Inference}},
	issn = {1556-5068},
	url = {https://www.ssrn.com/abstract=2676160},
	doi = {10.2139/ssrn.2676160},
	language = {en},
	urldate = {2020-08-05},
	journal = {SSRN Electronic Journal},
	author = {D'Orazio, Vito and Honaker, James and King, Gary},
	year = {2015}
}

@book{jungherrAnalyzingPoliticalCommunication2015,
	series = {Contributions to {Political} {Science}},
	title = {Analyzing {Political} {Communication} with {Digital} {Trace} {Data}: {The} {Role} of {Twitter} {Messages} in {Social} {Science} {Research}},
	isbn = {978-3-319-20318-8},
	shorttitle = {Analyzing {Political} {Communication} with {Digital} {Trace} {Data}},
	url = {https://www.springer.com/gp/book/9783319203188},
	language = {en},
	urldate = {2020-08-04},
	publisher = {Springer International Publishing},
	author = {Jungherr, Andreas},
	year = {2015},
	doi = {10.1007/978-3-319-20319-5}
}

@article{blondelSurveyResultsMobile2015,
	title = {A survey of results on mobile phone datasets analysis},
	volume = {4},
	copyright = {2015 Blondel et al.},
	issn = {2193-1127},
	url = {https://epjdatascience.springeropen.com/articles/10.1140/epjds/s13688-015-0046-0},
	doi = {10.1140/epjds/s13688-015-0046-0},
	language = {en},
	number = {1},
	urldate = {2020-08-04},
	journal = {EPJ Data Science},
	author = {Blondel, Vincent D. and Decuyper, Adeline and Krings, Gautier},
	month = dec,
	year = {2015},
	note = {Number: 1
Publisher: SpringerOpen},
	pages = {1--55}
}

@article{bolsenInfluencePartisanMotivated2014,
	title = {The {Influence} of {Partisan} {Motivated} {Reasoning} on {Public} {Opinion}},
	volume = {36},
	doi = {10.1007/s11109-013-9238-0},
	language = {en},
	number = {2},
	journal = {Political Behavior},
	author = {Bolsen, Toby and Druckman, James N. and Cook, Fay Lomax},
	month = jun,
	year = {2014},
	pages = {235--262},
}

@book{colemanFoundationsSocialTheory2000,
	address = {Cambridge, Mass.},
	title = {Foundations of social theory},
	isbn = {978-0-674-31226-5},
	language = {eng},
	publisher = {Belknap Press of Harvard Univ. Press},
	author = {Coleman, James Samuel},
	year = {1990}
}

@article{valkenburg2011gender,
  title={Gender differences in online and offline self-disclosure in pre-adolescence and adolescence},
  author={Valkenburg, Patti M and Sumter, Sindy R and Peter, Jochen},
  journal={British Journal of Developmental Psychology},
  volume={29},
  number={2},
  pages={253--269},
  year={2011},
  publisher={Wiley Online Library}
}

@article{van2019urban,
  title={Urban \& online: Social media use among adolescents and sense of belonging to a super-diverse city},
  author={van Eldik, Anne and Kneer, Julia and Jansz, Jeroen},
  journal={Media and Communication},
  volume={7},
  number={2},
  pages={242--253},
  year={2019}
}

@article{ahvanooey2020survey,
  title={A survey on smartphones security: Software vulnerabilities, malware, and attacks},
  author={Ahvanooey, Milad Taleby and Li, Qianmu and Rabbani, Mahdi and Rajput, Ahmed Raza},
  journal={arXiv preprint arXiv:2001.09406},
  year={2020}
}

@article{kreuter2016framing,
  title={The framing of the record linkage consent question},
  author={Kreuter, Frauke and Sakshaug, Joseph W and Tourangeau, Roger},
  journal={International Journal of Public Opinion Research},
  volume={28},
  number={1},
  pages={142--152},
  year={2016},
  publisher={Oxford University Press}
}

@book{pentland2010honest,
  title={Honest signals: how they shape our world},
  author={Pentland, Alex},
  year={2010},
  publisher={MIT press}
}

@article{perriam2020digital,
  title={Digital methods in a post-API environment},
  author={Perriam, Jessamy and Birkbak, Andreas and Freeman, Andy},
  journal={International Journal of Social Research Methodology},
  volume={23},
  number={3},
  pages={277--290},
  year={2020},
  publisher={Taylor \& Francis}
}

@article{pfeffer2018tampering,
  title={Tampering with Twitter’s sample API},
  author={Pfeffer, J{\"u}rgen and Mayer, Katja and Morstatter, Fred},
  journal={EPJ Data Science},
  volume={7},
  number={1},
  pages={50},
  year={2018},
  publisher={Springer Berlin Heidelberg}
}

@article{quan2010uses,
  title={Uses and gratifications of social media: A comparison of Facebook and instant messaging},
  author={Quan-Haase, Anabel and Young, Alyson L},
  journal={Bulletin of science, technology \& society},
  volume={30},
  number={5},
  pages={350--361},
  year={2010},
  publisher={SAGE Publications Sage CA: Los Angeles, CA}
}

@inproceedings{de2013online,
  title={Online data, fixed effects and the construction of high-frequency price indexes},
  author={de Haan, Jan and Hendriks, Rens},
  booktitle={Economic Measurement Group Workshop},
  pages={28--29},
  year={2013}
}

@article{singhTechnicalLookIndian2020,
	title = {A {Technical} {Look} {At} {The} {Indian} {Personal} {Data} {Protection} {Bill}},
	url = {http://arxiv.org/abs/2005.13812},
	urldate = {2020-08-12},
	journal = {arXiv:2005.13812 [cs]},
	author = {Singh, Ram Govind and Ruj, Sushmita},
	month = may,
	year = {2020},
	note = {arXiv: 2005.13812}
}

@article{sudaJapanPersonalInformation2020,
	title = {Japan’s {Personal} {Information} {Protection} {Policy} {Under} {Pressure}},
	volume = {60},
	issn = {0004-4687, 1533-838X},
	url = {https://online.ucpress.edu/as/article/60/3/510/110500/Japans-Personal-Information-Protection-Policy},
	doi = {10.1525/as.2020.60.3.510},
	language = {en},
	number = {3},
	urldate = {2020-08-12},
	journal = {Asian Survey},
	author = {Suda, Yuko},
	month = jun,
	year = {2020},
	pages = {510--533}
}

@article{wachterCounterfactualExplanationsOpening2017,
	title = {Counterfactual {Explanations} without {Opening} the {Black} {Box}: {Automated} {Decisions} and the {GDPR}},
	volume = {31},
	shorttitle = {Counterfactual {Explanations} without {Opening} the {Black} {Box}},
	url = {https://heinonline.org/HOL/P?h=hein.journals/hjlt31&i=860},
	language = {eng},
	number = {2},
	urldate = {2020-08-07},
	journal = {Harvard Journal of Law \& Technology (Harvard JOLT)},
	author = {Wachter, Sandra and Mittelstadt, Brent and Russell, Chris},
	year = {2017},
	pages = {841--888}
}

@article{gdpr2016,
  title   = {{Regulation (EU) 2016/679 of the European Parliament and of the Council of 27 April 2016 on the Protection of Natural Persons with Regard to the Processing of Personal Data and on the Free Movement of Such Data, and Repealing Directive 95/46/EC (General Data Protection Regulation)}},
  author = {{European Union}},
  journal={OJ},
  pages={1-89},
  volume={59 (L 119)},
  url={},
  year    = {2016}
}

@article{beyensEffectSocialMedia2020,
	title = {The effect of social media on well-being differs from adolescent to adolescent},
	volume = {10},
	issn = {2045-2322},
	url = {http://www.nature.com/articles/s41598-020-67727-7},
	doi = {10.1038/s41598-020-67727-7},
	language = {en},
	number = {1},
	urldate = {2020-08-07},
	journal = {Scientific Reports},
	author = {Beyens, Ine and Pouwels, J. Loes and van Driel, Irene I. and Keijsers, Loes and Valkenburg, Patti M.},
	month = dec,
	year = {2020},
	pages = {10763}
}

@article{dyreson1993timestamp,
  title={Timestamp semantics and representation},
  author={Dyreson, Curtis E and Snodgrass, Richard T},
  journal={Information Systems},
  volume={18},
  number={3},
  pages={143--166},
  year={1993},
  publisher={Elsevier}
}

@article{amayaTotalErrorBig2020,
	title = {Total {Error} in a {Big} {Data} {World}: {Adapting} the {TSE} {Framework} to {Big} {Data}},
	volume = {8},
	issn = {2325-0984, 2325-0992},
	shorttitle = {Total {Error} in a {Big} {Data} {World}},
	url = {https://academic.oup.com/jssam/article/8/1/89/5728725},
	doi = {10.1093/jssam/smz056},
	language = {en},
	number = {1},
	urldate = {2020-08-07},
	journal = {Journal of Survey Statistics and Methodology},
	author = {Amaya, Ashley and Biemer, Paul P. and Kinyon, David},
	month = feb,
	year = {2020},
	pages = {89--119}
}

@incollection{Biemer2016errors,
  author      = "Biemer, Paul P.",
  title       = "Errors and Inference",
  editor      = "Foster, Ian and Ghani, Rayid and Jarmin, Ron S and Kreuter, Frauke and Lane, Julia",
  booktitle   = "Big data and social science: A practical guide to methods and tools",
  publisher   = "CRC press",
  year        = 2016,
  pages       = "266-297",
  chapter     = 10,
}

@article{japec2015big,
  title={Big data in survey research: {AAPOR} task force report},
  author={Japec, Lilli and Kreuter, Frauke and Berg, Marcus and Biemer, Paul P. and Decker, Paul and Lampe, Cliff and Lane, Julia and O’Neil, Cathy and Usher, Abe},
  journal={Public Opinion Quarterly},
  volume={79},
  number={4},
  pages={839--880},
  year={2015},
  publisher={Oxford University Press US}
}

@article{bruns2019after,
  title={After the ‘APIcalypse’: social media platforms and their fight against critical scholarly research},
  author={Bruns, Axel},
  journal={Information, Communication \& Society},
  volume={22},
  number={11},
  pages={1544--1566},
  year={2019},
  publisher={Taylor \& Francis}
}

@book{bethlehem2011handbook,
  title={Handbook of nonresponse in household surveys},
  author={Bethlehem, Jelke and Cobben, Fannie and Schouten, Barry},
  volume={568},
  year={2011},
  publisher={John Wiley \& Sons}
}

@article{schafer2002missing,
  title={Missing data: our view of the state of the art.},
  author={Schafer, Joseph L and Graham, John W},
  journal={Psychological methods},
  volume={7},
  number={2},
  pages={147},
  year={2002},
  publisher={American Psychological Association}
}

@article{boeschoten2017obtain,
  title={How to Obtain Valid Inference under Unit Nonresponse?},
  author={Boeschoten, Laura and Vink, Gerko and Hox, Joop JCM},
  journal={Journal of Official Statistics},
  volume={33},
  number={4},
  pages={963--978},
  year={2017},
  publisher={Sciendo}
}

@article{shirima2007use,
  title={The use of personal digital assistants for data entry at the point of collection in a large household survey in southern Tanzania},
  author={Shirima, Kizito and Mukasa, Oscar and Schellenberg, Joanna Armstrong and Manzi, Fatuma and John, Davis and Mushi, Adiel and Mrisho, Mwifadhi and Tanner, Marcel and Mshinda, Hassan and Schellenberg, David},
  journal={Emerging themes in epidemiology},
  volume={4},
  number={1},
  pages={1--8},
  year={2007},
  publisher={BioMed Central}
}

@book{harron2015methodological,
  title={Methodological developments in data linkage},
  author={Harron, Katie and Goldstein, Harvey and Dibben, Chris},
  year={2015},
  publisher={John Wiley \& Sons}
}

@conference{scheermansecure2020,
  title={Secure Processing of Sensitive Data on shared HPC systems},
  author={Scheerman, M and Voort, L and Zarrabi, N},
  conference={CompBioMed2020},
  url={https://www.compbiomed-conference.org/wp-content/uploads/2020/03/Fri_11_15_Narges_Zarrabi-slides-compbiomed-website_smaller.pdf},
  year = {2020}
}

@article{patil2016statistical,
  title={A statistical definition for reproducibility and replicability},
  author={Patil, Prasad and Peng, Roger D and Leek, Jeffrey T},
  journal={BioRxiv},
  pages={066803},
  year={2016},
  publisher={Cold Spring Harbor Laboratory}
}

@misc{stier2019integrating,
  title={Integrating survey data and digital trace data: key issues in developing an emerging field},
  author={Stier, Sebastian and Breuer, Johannes and Siegers, Pascal and Thorson, Kjerstin},
  year={2019},
  publisher={SAGE Publications Sage CA: Los Angeles, CA}
}

@article{gayo2012no,
  title={No, you cannot predict elections with {Twitter}},
  author={Gayo-Avello, Daniel},
  journal={{IEEE} Internet Computing},
  volume={16},
  number={6},
  pages={91--94},
  year={2012},
  publisher={{IEEE}}
}

@inproceedings{dwork2008differential,
  title={Differential privacy: A survey of results},
  author={Dwork, Cynthia},
  booktitle={International conference on theory and applications of models of computation},
  pages={1--19},
  year={2008},
  organization={Springer}
}

@misc{UURDMS,
  title = {Utrecht University Research Data Management Support},
  howpublished = {\url{https://www.uu.nl/en/research/research-data-management}},
  note = {Accessed: 2020-07-18}
}

@article{wilkinson2016fair,
  title={The FAIR Guiding Principles for scientific data management and stewardship},
  author={Wilkinson, Mark D and Dumontier, Michel and Aalbersberg, IJsbrand Jan and Appleton, Gabrielle and Axton, Myles and Baak, Arie and Blomberg, Niklas and Boiten, Jan-Willem and da Silva Santos, Luiz Bonino and Bourne, Philip E and others},
  journal={Scientific data},
  volume={3},
  number={1},
  pages={1--9},
  year={2016},
  publisher={Nature Publishing Group}
}

@article{oberski2017evaluating,
  title={Evaluating the quality of survey and administrative data with generalized multitrait-multimethod models},
  author={Oberski, Daniel L and Kirchner, Antje and Eckman, Stephanie and Kreuter, Frauke},
  journal={Journal of the American Statistical Association},
  volume={112},
  number={520},
  pages={1477--1489},
  year={2017},
  publisher={Taylor \& Francis}
}

@article{manikonda2014analyzing,
  title={Analyzing user activities, demographics, social network structure and user-generated content on Instagram},
  author={Manikonda, Lydia and Hu, Yuheng and Kambhampati, Subbarao},
  journal={arXiv preprint arXiv:1410.8099},
  year={2014}
}

@article{ruktanonchai2018using,
  title={Using Google Location History data to quantify fine-scale human mobility},
  author={Ruktanonchai, Nick Warren and Ruktanonchai, Corrine Warren and Floyd, Jessica Rhona and Tatem, Andrew J},
  journal={International Journal of Health Geographics},
  volume={17},
  number={1},
  pages={28},
  year={2018},
  publisher={Springer}
}

@article{kramer2014experimental,
  title={Experimental evidence of massive-scale emotional contagion through social networks},
  author={Kramer, Adam DI and Guillory, Jamie E and Hancock, Jeffrey T},
  journal={Proceedings of the National Academy of Sciences},
  volume={111},
  number={24},
  pages={8788--8790},
  year={2014},
  publisher={National Acad Sciences}
}

@article{bouko2020emotions,
  title={Emotions through texts and images: A multimodal analysis of reactions to the Brexit vote on Flickr},
  author={Bouko, Catherine},
  journal={Pragmatics},
  volume={30},
  number={2},
  pages={222--246},
  year={2020},
  publisher={International Pragmatics Association (IPrA)}
}

@inproceedings{andrei2020signal,
  title={Signal performance analysis of the latest quartet of Galileo satellites during the first operational year},
  author={Andrei, Constantin-Octavian and Johansson, Jan and Koivula, Hannu and Poutanen, Markku},
  booktitle={2020 International Conference on Localization and GNSS (ICL-GNSS)},
  pages={1--6},
  year={2020},
  organization={IEEE}
}

@article{brakenhoff2018measurement,
  title={Measurement error is often neglected in medical literature: a systematic review},
  author={Brakenhoff, Timo B and Mitroiu, Marian and Keogh, Ruth H and Moons, Karel GM and Groenwold, Rolf HH and van Smeden, Maarten},
  journal={Journal of clinical epidemiology},
  volume={98},
  pages={89--97},
  year={2018},
  publisher={Elsevier}
}

@article{bland1996measurement,
  title={Measurement error and correlation coefficients.},
  author={Bland, J Martin and Altman, Douglas G},
  journal={BMJ: British Medical Journal},
  volume={313},
  number={7048},
  pages={41},
  year={1996},
  publisher={BMJ Publishing Group}
}

@book{biemer2011latent,
  title={Latent class analysis of survey error},
  author={Biemer, Paul P.},
  volume={571},
  year={2011},
  publisher={John Wiley \& Sons}
}

@book{carroll2006measurement,
  title={Measurement error in nonlinear models: a modern perspective},
  author={Carroll, Raymond J and Ruppert, David and Stefanski, Leonard A and Crainiceanu, Ciprian M},
  year={2006},
  publisher={CRC press}
}

@article{hsu2002face,
  title={Face detection in color images},
  author={Hsu, Rein-Lien and Abdel-Mottaleb, Mohamed and Jain, Anil K},
  journal={IEEE transactions on pattern analysis and machine intelligence},
  volume={24},
  number={5},
  pages={696--706},
  year={2002},
  publisher={IEEE}
}

@article{hjelmaas2001face,
  title={Face detection: A survey},
  author={Hjelmaas, Erik and Low, Boon Kee},
  journal={Computer vision and image understanding},
  volume={83},
  number={3},
  pages={236--274},
  year={2001},
  publisher={Elsevier}
}

@article{kaya2017video,
  title={Video-based emotion recognition in the wild using deep transfer learning and score fusion},
  author={Kaya, Heysem and G{\"u}rpinar, Furkan and Salah, Albert Ali},
  journal={Image and Vision Computing},
  volume={65},
  pages={66--75},
  year={2017},
  publisher={Elsevier}
}

@article{lohr2008coverage,
  title={Coverage and sampling},
  author={Lohr, Sharon L},
  journal={International handbook of survey methodology},
  pages={97--112},
  year={2008},
  publisher={Lawrence Erlbaum Associates New York}
}

@incollection{lohr2009multiple,
  title={Multiple-frame surveys},
  author={Lohr, Sharon L},
  booktitle={Handbook of statistics},
  volume={29},
  pages={71--88},
  year={2009},
  publisher={Elsevier}
}

@book{cochran2007sampling,
  title={Sampling techniques},
  author={Cochran, William G},
  year={2007},
  publisher={John Wiley \& Sons}
}

@book{de2008international,
  title={International handbook of survey methodology.},
  author={De Leeuw, Edith D and Hox, Joop J and Dillman, Don A},
  year={2008},
  publisher={Taylor \& Francis Group/Lawrence Erlbaum Associates}
}

@book{chambers2003analysis,
  title={Analysis of survey data},
  author={Chambers, Ray L and Skinner, Chris J},
  year={2003},
  publisher={John Wiley \& Sons}
}

@article{groves2008impact,
  title={The impact of nonresponse rates on nonresponse bias: a meta-analysis},
  author={Groves, Robert M and Peytcheva, Emilia},
  journal={Public opinion quarterly},
  volume={72},
  number={2},
  pages={167--189},
  year={2008},
  publisher={Oxford University Press}
}

@article{singer1993informed,
  title={Informed consent and survey response: a summary of the empirical literature},
  author={Singer, Eleanor},
  journal={Journal of Official Statistics},
  volume={9},
  number={2},
  pages={361},
  year={1993},
  publisher={Statistics Sweden (SCB)}
}

@article{zhang2012topics,
  title={Topics of statistical theory for register-based statistics and data integration},
  author={Zhang, Li-Chun},
  journal={Statistica Neerlandica},
  volume={66},
  number={1},
  pages={41--63},
  year={2012},
  publisher={Wiley Online Library}
}

@article{doidge2019reflections,
  title={Reflections on modern methods: linkage error bias},
  author={Doidge, James C and Harron, Katie L},
  journal={International Journal of Epidemiology},
  volume={48},
  number={6},
  pages={2050--2060},
  year={2019},
  publisher={Oxford University Press}
}

@article{kim2020data,
  title={Data integration by combining big data and survey sample data for finite population inference},
  author={Kim, Jae-kwang and Tam, Siu - Ming},
  journal={arXiv preprint arXiv:2003.12156},
  year={2020}
}

@book{van2020general,
  title={The General Data Protection Regulation in Plain Language},
  author={van der Sloot, Bart},
  year={2020},
  publisher={Amsterdam University Press}
}

@book{privitera2018research,
  title={Research methods for the behavioral sciences},
  author={Privitera, Gregory J},
  year={2018},
  publisher={Sage Publications}
}

@article{wong2019right,
  title={The right to data portability in practice: exploring the implications of the technologically neutral GDPR},
  author={Wong, Janis and Henderson, Tristan},
  journal={International Data Privacy Law},
  volume={9},
  number={3},
  pages={173--191},
  year={2019},
  publisher={Oxford University Press}
}

@article{ausloos2019getting,
  title={Getting Data Subject Rights Right: A submission to the European Data Protection Board from international data rights academics, to inform regulatory guidance},
  author={Ausloos, J and Veale, M and Mahieu, R and others},
  journal={Journal of Intellectual Property, Information Technology and Electronic Commerce Law},
  volume={10},
  year={2019}
}

@article{boeschoten2018updating,
  title={Updating latent class imputations with external auxiliary variables},
  author={Boeschoten, Laura and Oberski, Daniel L and De Waal, Ton and Vermunt, Jeroen K},
  journal={Structural Equation Modeling: A Multidisciplinary Journal},
  volume={25},
  number={5},
  pages={750--761},
  year={2018},
  publisher={Taylor \& Francis}
}

@article{guerra2010occupants,
  title={Occupants' behaviour: determinants and effects on residential heating consumption},
  author={Guerra-Santin, Olivia and Itard, Laure},
  journal={Building Research \& Information},
  volume={38},
  number={3},
  pages={318--338},
  year={2010},
  publisher={Taylor \& Francis}
}

@article{breedveld2002background,
  title={Background to the methods used in the Time Budget Survey (TBO)},
  author={Breedveld, Koen and Van Den Broek, Andries and Huysmans, Frank},
  journal={Social and Cultural Planning Office of the Netherlands. Disponible en internet via: http://www. scp. nl/onderzoek/tbo/english/achtergronden/history. pdf},
  year={2002},
  publisher={Citeseer}
}

@book{papacharissi2010networked,
  title={A networked self: Identity, community, and culture on social network sites},
  author={Papacharissi, Zizi},
  year={2010},
  publisher={Routledge}
}

@article{freelon2018computational,
  title={Computational research in the post-API age},
  author={Freelon, Deen},
  journal={Political Communication},
  volume={35},
  number={4},
  pages={665--668},
  year={2018},
  publisher={Taylor \& Francis}
}

@article{andrews2015beyond,
  title={Beyond self-report: tools to compare estimated and real-world smartphone use},
  author={Andrews, Sally and Ellis, David A and Shaw, Heather and Piwek, Lukasz},
  journal={PloS one},
  volume={10},
  number={10},
  pages={e0139004},
  year={2015},
  publisher={Public Library of Science}
}

@article{reeves2019screenomics,
  title={Screenomics: A framework to capture and analyze personal life experiences and the ways that technology shapes them},
  author={Reeves, Byron and Ram, Nilam and Robinson, Thomas N and Cummings, James J and Giles, C Lee and Pan, Jennifer and Chiatti, Agnese and Cho, MJ and Roehrick, Katie and Yang, Xiao and others},
  journal={Human--Computer Interaction},
  pages={1--52},
  year={2019},
  publisher={Taylor \& Francis}
}

@article{haenschen2020self,
  title={Self-reported versus digitally recorded: Measuring political activity on Facebook},
  author={Haenschen, Katherine},
  journal={Social Science Computer Review},
  volume={38},
  number={5},
  pages={567--583},
  year={2020},
  publisher={SAGE Publications Sage CA: Los Angeles, CA}
}

@article{scharkow2019reliability,
  title={The reliability and temporal stability of self-reported media exposure: A meta-analysis},
  author={Scharkow, Michael},
  journal={Communication Methods and Measures},
  volume={13},
  number={3},
  pages={198--211},
  year={2019},
  publisher={Taylor \& Francis}
}

@article{araujo2017much,
  title={How much time do you spend online? Understanding and improving the accuracy of self-reported measures of Internet use},
  author={Araujo, Theo and Wonneberger, Anke and Neijens, Peter and de Vreese, Claes},
  journal={Communication Methods and Measures},
  volume={11},
  number={3},
  pages={173--190},
  year={2017},
  publisher={Taylor \& Francis}
}

@article{halavais2019overcoming,
  title={Overcoming terms of service: a proposal for ethical distributed research},
  author={Halavais, Alexander},
  journal={Information, Communication \& Society},
  volume={22},
  number={11},
  pages={1567--1581},
  year={2019},
  publisher={Taylor \& Francis}
}

@inproceedings{buolamwini2018gender,
  title={Gender shades: Intersectional accuracy disparities in commercial gender classification},
  author={Buolamwini, Joy and Gebru, Timnit},
  booktitle={Conference on fairness, accountability and transparency},
  pages={77--91},
  year={2018}
}

@article{lovestone2020european,
  title={The European medical information framework: A novel ecosystem for sharing healthcare data across Europe},
  author={Lovestone, Simon and EMIF Consortium},
  journal={Learning Health Systems},
  volume={4},
  number={2},
  pages={e10214},
  year={2020},
  publisher={Wiley Online Library}
}

@article{sen2019total,
  title={A total error framework for digital traces of humans},
  author={Sen, Indira and Floeck, Fabian and Weller, Katrin and Weiss, Bernd and Wagner, Claudia},
  journal={arXiv preprint arXiv:1907.08228},
  year={2019}
}

@article{Beinhauer2020,
  title={Towards a total error framework for sensor and survey data},
  author={Beinhauer, Lukas and Snijkers, Ger and Bakker, Jeldrik},
  journal={BigSurv20},
  year={2020}
}

\end{document}